\documentclass[aps,pre,twocolumn,showpacs,superscriptaddress,groupedaddress]{revtex4}  
\usepackage{amsfonts}
\usepackage{amsmath}
\usepackage{amssymb}
\usepackage{version}
\usepackage{color}
\usepackage{graphicx}%
\includeversion{New_connection}
\excludeversion{Old_connection}

\begin{document}
\title{Langevin dynamics in inhomogeneous media: Re-examining the It\^{o}-Stratonovich Dilemma}
\author{Oded Farago}
\affiliation{Department of Mechanical and Aerospace Engineering, University of California, Davis, CA 95616}
\affiliation{Department of Biomedical Engineering, Ben Gurion University of the Negev, Be'er Sheva, 84105 Israel}
\affiliation{Ilse Katz Institute for Nanoscale Science and Technology, Ben Gurion University of the Negev, Be'er Sheva, 84105 Israel}
\author{Niels Gr{\o}nbech-Jensen}
\affiliation{Department of Mechanical and Aerospace Engineering, University of California, Davis, CA 95616}
\affiliation{Department of Chemical Engineering and Materials Science, University of California, Davis, CA 95616}
\affiliation{Computational Research Division, Lawrence Berkeley National Laboratory, Berkeley, CA 94720}

\begin{abstract}

The diffusive dynamics of a particle in a medium with space-dependent
friction coefficient is studied within the framework of the inertial
Langevin equation. In this description, the ambiguous interpretation
of the stochastic integral, known as the It\^{o}-Stratonovich dilemma,
is avoided since all interpretations converge to the same solution in
the limit of small time steps. We use a newly developed method for
Langevin simulations to measure the probability distribution of a
particle diffusing in a flat potential. Our results reveal that both
the It\^{o} and Stratonovich interpretations converge very slowly to
the uniform equilibrium distribution for vanishing time step
sizes. Three other conventions exhibit significantly improved
accuracy: (i) the ``isothermal'' (H\"{a}nggi) convention, (ii) the
Stratonovich convention corrected by a drift term, and (iii) a new
convention employing two different effective friction coefficients
representing two different averages of the friction function during
the time step.  We argue that the most physically accurate dynamical
description is provided by the third convention, in which the particle
experiences a drift originating from the dissipation instead of the
fluctuation term. This feature is directly related to the fact that
the drift is a result of an inertial effect that cannot be well
understood in the Brownian, overdamped limit of the Langevin equation.

\end{abstract}
\maketitle

\section{Introduction}
\label{intro}

The dynamics of a particle coupled to a thermal heat bath is
traditionally described by the Langevin equation of motion.  Denoting
the position and velocity of the particle at time $t$ by $r(t)$ and
$v(t)=\dot{r}$, respectively, the Langevin equation reads
\cite{coffey}
\begin{eqnarray}
m\dot{v}&=&f(r,t)-\alpha v+\beta(t),
\label{Lang_EOM}
\end{eqnarray}
where $m$ is the mass of the particle. This is Newtonian dynamics for
which the applied force is described as a composite of three kinds:
(i) deterministic forces $f(r,t)$, (ii) a friction force $-\alpha v$
proportional to the velocity with friction coefficient $\alpha\ge0$,
and (iii) a stochastic force $\beta(t)$ representing fluctuations
arising from interactions with the embedding medium that produces the
friction. The stochastic force can be conveniently modeled by a
delta-correlated (``white'') Gaussian noise with statistical
properties: $\langle \beta(t)\rangle=0$ and $\langle
\beta(t)\beta(t^\prime) \rangle = 2\alpha k_BT\delta(t-t^\prime)$,
where $k_B$ is Boltzmann's constant and $T$ is the thermodynamic
temperature \cite{parisi}. This definitions ensure that, over large
times, the position of the particle produces the Boltzmann
distribution $\rho(r)\sim \exp[-U(r)/k_BT]$, where $f=-\nabla U(r)$,
$U(r)$ being the potential energy surface \cite{kampen}. Moreover, in
the case of a flat potential (i.e., for $f=0$), the particle exhibits
diffusive behavior with diffusion coefficient $D=k_BT/\alpha$ - in
agreement with the fluctuation-dissipation theorem \cite{kubo}

Implied in the above description is that the friction coefficient
$\alpha$ is a constant. Now consider the particle diffusing in an
inhomogeneous medium with position-dependent friction coefficient
$\alpha(r)$. Presumably, the dynamics of a particle in such an
environment is governed by Eq.~(\ref{Lang_EOM}), with the constant
$\alpha$ simply replaced by the position-dependent $\alpha(r)$. This
is formally correct, but physically meaningless, as readily becomes
obvious when one attempts to calculate the particle trajectory by
integrating Eq.~(\ref{Lang_EOM}) over a short time interval $dt$
\cite{doob}. For a uniform $\alpha$, the integral over the stochastic
noise is a well-defined Wiener process
\begin{eqnarray}
\int_0^{dt} \beta(t^\prime) \, dt^\prime =\sqrt{2\alpha k_BT dt}\, \sigma,
\label{eq:weiner}
\end{eqnarray}
where $\sigma$ is a standard Gaussian random number with
$\langle\sigma\rangle=0$ and $\langle\sigma^2\rangle=1$
\cite{coffey,kampen}. For a non-uniform $\alpha(r)$, the integral is
ill-defined, since one needs to specify at which point along the
trajectory the friction coefficient in Eq.~(\ref{eq:weiner}) is
evaluated. This ambiguity is known as the {\em It\^{o}-Stratonovich
dilemma}\/ \cite{coffey,mannella}. In It\^{o}'s interpretation
\cite{ito}, the friction coefficient is taken at the beginning of the
time step, while the Stratonovich interpretation considers the
algebraic mean of the initial and final frictions \cite{stratonovich}
(which, for a small time interval, is very close to the friction at
the mid-point). In the Brownian (overdamped) limit of the Langevin
equation (i.e., when the left hand side of Eq.~(\ref{Lang_EOM})
vanishes), neither of these two interpretations reproduce the Boltzmann
distribution. Instead, it is the the so called ``isothermal''
(H\"{a}nggi) convention \cite{hanggi}, which takes the friction
coefficient at the end of the time step, that generates the correct
equilibrium statistics \cite{lau,footnote}.

Equilibrium statistics may also be acheived within the It\^{o} or the
Stratonovich formulation for Brownian dynamics. In these conventions,
however, the associated Focker-Planck equation must be augmented with
a ``spurious'' force term proportional to the gradient of the
logarithm of the friction coefficient
$f_s\sim-k_BT\alpha^\prime(r)/\alpha(r)$ \cite{lau,volpe,sancho} (see
further discussion in section \ref{sec:conclusions} on the ``corrected
Stratonovich convention'' employing a spurious drift correction
associated with this extra term). The necessity of such spurious force
term can be understood through the following example
\cite{tupper}. Consider a Brownian particle diffusing between two
immiscible fluids with a sharp interface between them. In the absence
of an external potential, the equilibrium distribution is uniform, and
the particle is expected, on average, to spend an equal amount time on
both sides of the interface. However, this expectation appears to
contradict the observation that the particle diffuses slower, and
therefore becomes trapped, on the more viscous side. These conflicting
arguments can be reconciled by assuming that there exists a
boundary-force that pushes the particle toward the less viscous
fluid. Thus, the particle has a smaller probability to enter the more
viscous side in a manner that exactly cancels the trapping effect.

\begin{figure}[t]
\begin{center}
\scalebox{0.525}{\centering \includegraphics{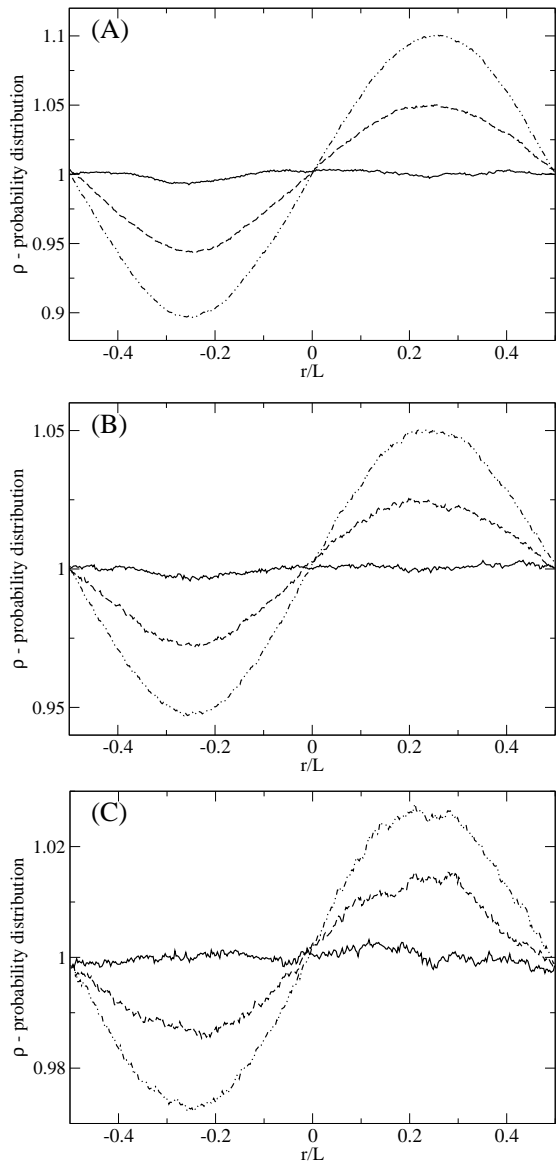}}
\end{center}
\vspace{-0.5cm}
\caption{Probability distribution computed using It\^{o}
(dashed-dotted), Stratonovich (dashed), isothermal (solid line)
conventions, for (A) $dt=0.1$, (B) $dt=0.05$, and (C) $dt=0.025$.
Trajectories were generated using
Eqs.~(\ref{pos-equation})-(\ref{vel-equation}), with
$\bar{\alpha}_r=\bar{\alpha}_t=\alpha(r^n)$ (It\^{o}),
$\bar{\alpha}_r=\bar{\alpha}_t=\int_{r^n}^{r^{n+1}}\!\!\!\!\alpha(r^\prime)dr^\prime/(r^{n+1}-r^n)$
(Stratonovich), $\bar{\alpha}_r=\bar{\alpha}_t=\alpha(r^{n+1})$
(isothermal).}
\label{fig:results}
\end{figure}

So far there has been relatively little discussion of the
It\^{o}-Stratonovich dilemma in the framework of the full inertial
Langevin equation. (Recent noticeable exceptions includes the
treatment in Refs.~\cite{sancho,kupferman} where the Brownian
overdamped limit has been approached through ``adiabatic elimination''
of the inertial degrees of freedom.) This can be attributed to the
less ``catastrophic'' consequences of the dilemma on the second order
differential Langevin equation in comparison to the first order
overdamped Brownian equation of motion. Suppose $r(t)$ is calculated
by numerically integrating the equation of motion with $N$ small time
steps of size $dt=t/N$. For the first order Brownian equation the
velocity diverges, resulting in {\it different} results of $r(t)$ when
applying different interpretations (It\^{o}, Stratonovich,
isothermal); even for $dt\rightarrow0$ (see Ref.~\cite{risken},
section 3.3.3).  In other words, no matter how small the integration
time step is, different interpretations yield different trajectories,
and the method of integration therefore seems accurate only to zeroth
order in $dt$. In contrast, it is the acceleration that diverges for
the second order Langevin equation, while the velocity remains
finite. This means that all interpretations generate similar
trajectories in the limit of vanishing time step sizes. While this
definitely reduces the severity of the dilemma, it leaves open an
important practical question concerning the rate of convergence of
numerical integrators implementing different interpretations. The
problem is demonstrated in Fig.~\ref{fig:results}, showing the spatial
equilibrium distribution computed from Langevin dynamics simulations
of a particle of mass $m=1$, in contact with a constant temperature
bath $k_BT=1$, moving in a one-dimensional medium with a flat
potential and a sinusoidal friction coefficient given by
$\alpha(r)=2.75+2.25\sin(2\pi r/L)$, where $L=40$. The simulations
employ the efficient G-JF Langevin integrator \cite{gjf} (see also
Eqs.~(\ref{pos-equation})-(\ref{vel-equation}) below), which has
recently been introduced for simulations with constant $\alpha$. It is
here modified according to the way that the friction coefficient is
defined in each convention (see details in caption of
Fig.~\ref{fig:results}). Since the potential energy is constant, the
equilibrium distribution must be uniform. Our results in
Fig.~\ref{fig:results} show that both It\^{o} and Stratonovich
interpretations exhibit noticeable deviations from the correct uniform
equilibrium distribution, and the deviations reflect the sinusoidal
form of the friction function. Their magnitude seem to decrease
linearly with $dt$, where the Stratonovich convention is approximately
twice as accurate as It\^{o}'s for a given $dt$. In contrast, the
isothermal convention produces fairly uniform distributions (albeit
somewhat noisy) which, even for $dt=0.1$, only deviates by less than
$0.5\%$ from the correct value of 1.

We here argue that the success of the isothermal convention stems from
the cancellation of two errors that it makes in the evaluations of the
friction and noise terms in the Langevin equation of motion. We
further suggest that there exists an alternative formulation that also
generates the correct distribution for large time steps without making
the errors in the dynamical description. This alternative convention
is closely related to the more physical Stratonovich convention in
which the friction coefficient is averaged over the path made by the
particle during a time step. The new scheme does {\em not}\/ require
the addition of a drift term to the Langevin equation of
motion. Instead, the drift originates from the friction term, not the
noise, in the Langevin equation. We conclude that the spurious drift
is an {\em inertial effect}\/.

\section{The New Convention}
\label{newconvention}

We follow a similar route to the one we have used in our treatment of
homogeneous media with constant $\alpha$ \cite{gjf}, and integrate
Eq.~(\ref{Lang_EOM}) over a small time interval from $t_n$ to
$t_{n+1}=t_n+dt$. This yields
\begin{eqnarray}
m\left(v^{n+1}-v^n\right)=\int_{t_n}^{t_{n+1}}\!\!\!\!\! f
dt^\prime-\!\!\int_{r_n}^{r_{n+1}}\!\!\!\!\! \alpha (r^\prime)dr^\prime
+\!\!\int_{t_n}^{t_{n+1}}\!\!\!\!\!
\beta dt^\prime, \nonumber \\
\label{eq:verlet1}
\end{eqnarray}
where $r^n$ and $v^n$ represent the particle position and velocity at
$t=t_n$, respectively. Equation (\ref{eq:verlet1}) reveals the following
important property of the fluctuation-dissipation relation in
inhomogeneous systems. The integral of the second term in
the equation, representing the total change in momentum due
to the friction force, gives
\begin{eqnarray}
\int_{r_n}^{r_{n+1}}\!\!\!\!\! \alpha (r^\prime)dr^\prime=\bar{\alpha}_r
\left(r^{n+1}-r^n\right),
\label{eq:rfriction}
\end{eqnarray}
where $\bar{\alpha}_r$ is the {\em spatial average}\/ of the friction
coefficient along the interval that the particle has traveled. We note
that $\bar{\alpha}_r=[A(r^{n+1})-A(r^n)]/(r^{n+1}-r^n)$, where $A(r)$
is the primitive function of $\alpha(r)$. For smooth friction
functions, $\bar{\alpha}_r$ is closely related to the Stratonovich
friction coefficient, $\alpha_s$, namely
$\bar{\alpha}_r\simeq\alpha_{\rm
s}\simeq\alpha(r^n)+\alpha^\prime(r^n)(r^{n+1}-r^n)/2$. In contrast,
the meaning of the integral over the noise term in
Eq.~(\ref{eq:verlet1}) (last term on the r.h.s.) is less clear. Since
this is the sum of random Gaussian variables with vanishing
correlation time, it can be {\em formally}\/ written as an integral of
a Wiener process (compare with Eq.~(\ref{eq:weiner})):
\begin{eqnarray}
\int_{t_n}^{t_{n=1}} \beta(t^\prime) \,
dt^\prime =\sqrt{2\bar{\alpha}_t k_BT dt}\, \sigma,
\label{eq:tfriction}
\end{eqnarray}
where $\bar{\alpha}_t$ is the {\em temporal average}\/ of the friction
coefficient during the time step. A striking observation that can be
immediately made is that unlike the case of a constant friction, where
$\bar{\alpha}_r=\bar{\alpha}_t=\alpha$, in the nonuniform case
$\bar{\alpha}_r$ and $\bar{\alpha}_t$ are generally different.There is
a fundamental difference between $\bar{\alpha}_r$ and
$\bar{\alpha}_t$. While the former can be evaluated through
Eq.~(\ref{eq:rfriction}) from the values of the end-points $r^n$ and
$r^{n+1}$ (provided that the function $\alpha(r)$ is known), the
evaluation of the latter requires full knowledge of how the path
$r(t)$ is traveled during the time step. Without this information
$\bar{\alpha}_t$ cannot be uniquely determined for a given $r^n$ and
$r^{n+1}$, since there exist not only one path but an ensemble of
trajectories leading from $r^n$ to $r^{n+1}$. This is the origin of
the dilemma. While Eq.~(\ref{eq:tfriction}) is formally correct, it
has no unique physical meaning for finite time steps as it is based on
the assumption that the noise is temporally uncorrelated (white),
which is only true for infinitesimally small $dt$. Over finite time
steps, the friction gradient colors the noise, since the noise value
at one time instance changes the trajectory of the particle and,
thereby, influences the noise statistics at subsequent time instances.

\begin{figure}[t]
\begin{center}
\scalebox{0.525}{\centering \includegraphics{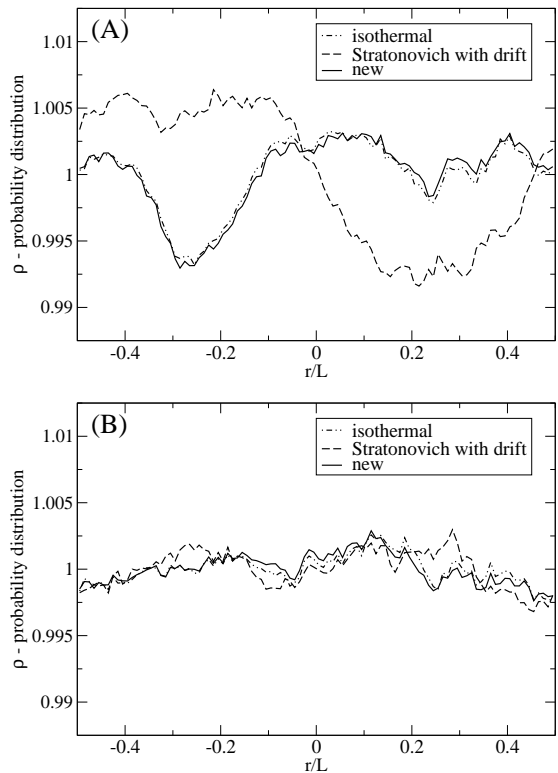}}
\end{center}
\vspace{-0.5cm}
\caption{Probability distribution computed using the isothermal
(dashed-dotted), corrected-Stratonovich (dashed), and the new
convention of this paper (solid line), for (A) $dt=0.1$ and (B)
$dt=0.025$. Results for the isothermal convention are replotted from
Fig.~\ref{fig:results}.  In the new convention, the trajectories are
generated using Eqs.~(\ref{pos-equation})-(\ref{vel-equation}), with
$\bar{\alpha}_r$ and $\bar{\alpha}_t$ given by Eq~(\ref{eq:rfriction})
and (\ref{eq:tfriction2}), respectively. In the corrected-Stratonovich
convention, both $\bar{\alpha}_r$ and $\bar{\alpha}_t$ are given by
Eq~(\ref{eq:rfriction}), and the newly computed position, $r^{n+1}$,
is shifted by $-[\alpha^\prime(r^n)/\alpha(r^{n})](k_BT/m)(dt^2/4)$ to
correct for the spurious drift.}
\label{fig:results2}
\end{figure}

Within the Stratonovich and isothermal conventions,
Eq.~(\ref{eq:tfriction}) is interpreted as if the impulse of the
stochastic force is a product of a standard Gaussian number $\sigma$
and a quantity $\sqrt{2\bar{\alpha}_t k_BT dt}$ that depends on the
particle's coordinates at both the beginning and end of the time
step. The dependency on the end-point implies that the mean stochastic
impulse does not vanish (to be shown later in section
\ref{sec:conclusions}). This feature, however, is problematic from the
physical view point of the fluctuation-dissipation theorem. An
insightful way to realize this is provided by Gillespie's
classical-mechanical derivation of the Langevin equation (see
Ref.~\cite{gillespie}, section 4.5). In this {\em ab-initio}\/
treatment, one considers the motion of a heavy particle in a bath of
lighter particles with which it collides elastically. Assuming that
the velocities of the bath particles are drawn from a
Maxwell-Boltzmann distribution, it can be shown that the collisions
produce a stochastic force whose mean is given by the friction force
in the differential Langevin equation (\ref{Lang_EOM}), while the
noise term accounts for the fluctuations around the mean value.  This
implies that, when one considers the average over all possible noise
realizations, the mean change in momentum due to the noise must
vanish, and this feature must be incorporated in the integral form of
the Langevin equation (\ref{eq:verlet1}) to make it consistent with
the fluctuation-dissipation relationship.  In other words, within the
framework of inertial Langevin dynamics, the noise term must be drawn
from a distribution with zero average value, which leads to two
seemingly conflicting conclusions. On the one hand, it suggests that
the noise should not depend on the end-point. On the other hand, the
noise is governed by the time-averaged friction coefficient
$\bar{\alpha}_t$, which makes in unclear how it cannot be dependent on
the particle's destination. This seeming paradox can be avoided by
determining $\bar{\alpha}_t$ based on the information existing at the
beginning of the time step, namely $r^n$ and $v^n$,and using the
a-priori best guess for the particle's trajectory. Since $v\simeq
v^n$, to leading order, we can assume that the particle travels with a
constant velocity $v^n$ during the time step, which yields the
approximation
\begin{eqnarray}
\bar{\alpha}_t\simeq \alpha(r^n)+\alpha^\prime(r^n)\frac{v^n dt}{2},
\label{eq:tfriction2}
\end{eqnarray}
to be used in Eq.~(\ref{eq:tfriction}). Combining this result
with Eq~(\ref{eq:rfriction}) for the position-averaged friction that
governs the dissipation term, we derive (in a manner similar to the
one outlined in Ref.~\cite{gjf}) the following velocity-explicit Verlet
Langevin integrator
\begin{eqnarray}
\!\!\!\!r^{n+1}\!&=&r\!^n+bdtv^n+\frac{bdt^2}{2m}f^n+\frac{bdt}{2m}\sqrt{2\bar{\alpha}_t
k_BTdt}\,\sigma^{n+1}\label{pos_verlet}\nonumber \\ 
\label{pos-equation}\\
\!\!\!\!v^{n+1}\!&=&\!av^n\!+\!\frac{dt}{2m}\left(af^n+f^{n+1}\right)\!+\!\frac{b}{m}\sqrt{2\bar{\alpha}_t
k_BTdt}\,\sigma^{n+1}\label{vel_verlet}, \nonumber \\
\label{vel-equation}
\end{eqnarray}
where $f^n=f(r^n)$, $\sigma^n$ is a random Gaussian number with zero
mean and unity variance, and the coefficients $a$ and $b$ are given
by: $b=(1+\bar{\alpha}_r dt/2m)^{-1}$, and $a=b(1-\bar{\alpha}_r
dt/2m)$.  Figure \ref{fig:results2} demonstrates that our new
convention, which assigns different values for the friction in the
fluctuation and dissipation terms, generates a probability
distribution that is essentially identical to the accurate
distribution of the isothermal interpretation. The figure, in fact,
reveals that when using the same seed for the random number generator,
these two conventions generate almost identical trajectories and,
thus, they reproduce probability distributions that are
indistinguishable of each other in fine details. This observation
presents a new puzzle since our new convention is closely related to
the Stratonovich convention that uses the position-averaged friction
in both terms and not to the isothermal convention that uses the
friction value at the end-point. The solution, discussed below,
provides physical insight into the so-called spurious drift generated
by the friction gradient.

\section{The Drift}

We have argued above that the noise-term in Eq.~(\ref{eq:verlet1})
should have zero mean when averaged over all possible noise
realizations during one time step. To satisfy the second law of
thermodynamics, the friction term in Eq.~(\ref{eq:verlet1}) must also
have a zero mean; yet, this is not true for each time step, but rather
upon averaging over an ensemble of Brownian particles that leave from
$r^n$ with the Maxwell-Boltzmann velocity distribution (or,
equivalently, upon averaging over all the occasions that a single
particle travels through at $r^n$ in the long time limit). This
requirement is necessary to ensure that the statistical-mechanical
average change in the momentum of the particle vanishes and,
therefore, that no real force acts on the particle while it diffuses
in a flat potential. Thus, $\langle\bar{\alpha}_r\Delta r\rangle=0$,
where $\Delta r=r^{n+1}-r^n$ ensures that no average momentum is
contributed to the particle by the noise. Using the approximation
$\bar{\alpha}_r\simeq \alpha(r^n)+\alpha^\prime (r^n)\Delta r/2$, this
condition leads to
\begin{eqnarray}
\alpha(r^n)\langle \Delta r\rangle \simeq -\frac{\alpha^\prime
(r^n)}{2}\left \langle (\Delta r)^2\right\rangle=-\frac{\alpha^\prime
(r^n)}{2}\frac{k_BT}{m}dt^2,
\label{eq:drift1}
\end{eqnarray}
where the second equality is obtained by virtue of the equilibrium
Maxwell-Boltzmann velocity distribution, and the fact that to linear
order in $dt$, $\Delta r\simeq v^n dt$ and, therefore, $\langle
(\Delta r)^2\rangle \simeq \langle (v^n)^2 \rangle
dt^2=(k_BT/m)dt^2$. From Eq.~(\ref{eq:drift1}) we conclude that the
particle's mean drift is given by
\begin{eqnarray}
\langle \Delta r\rangle = -\frac{1}{2}\frac{\alpha^\prime
(r^n)}{\alpha(r^n)} \frac{k_BT}{m}dt^2.
\label{eq:drift2}
\end{eqnarray}
The associated force $f_s=-k_BT\alpha^\prime/\alpha$, is the force
that produces the same mean drift in a uniform medium, in which case
$\langle \Delta r\rangle=(f/2m)dt^2$. This force is obviously not
real, which is why it is denoted ``spurious'' \cite{footnote2}.  Our
derivation demonstrates that the drift originates from the
dissipation, not the fluctuation, term in Langevin equation of
motion. This observation is directly related to the notion that the
drift represents an inertial effect arising from the feature that when
the particle travels toward a less viscous regime, it experiences less
friction and therefore travels longer distances. This physical picture
cannot be captured within the framework of overdamped Brownian
dynamics in which the inertial degree of freedom is
missing. Furthermore, our derivation suggests that another term
frequently used in the literature; namely ``noise induced drift'' is
somewhat misleading. This term stems from the proportionality of the
drift to $k_BT$. However, as our deviation demonstrates, the noise
term does not produce the drift. Instead, the linearity of $\langle
\Delta r\rangle$ with temperature is due to mean squared velocity
(which, per se, is obviously of thermal origin). At higher
temperatures, the particle moves faster and this amplifies the
magnitude of the inertial effect that produces the drift.

\section{Concluding Remarks}
\label{sec:conclusions}

We conclude by explaining why the seemingly more physical Stratonovich
interpretation fails to produce the correct equilibrium distribution,
and why the isothermal convention succeeds. We first note that by
multiplying Eq.~(\ref{pos_verlet}) by $\sigma^{n+1}$, and by using the
leading order approximations $\bar{\alpha}\simeq \alpha(r^n)$ (this is
true in any convention as long as $\alpha(r)$ is sufficiently smooth),
$b\simeq 1$, and the relations $\langle v^n\sigma^{n+1}\rangle =0$,
$\langle\sigma^{n+1}\rangle=0$, and $\langle
(\sigma^{n+1})^2\rangle=1$, we find that
\begin{eqnarray}
\langle \sigma^{n+1}\Delta r\rangle\simeq \sqrt{2\alpha k_BT
dt}\,(dt/2m),
\label{eq:sigmar}
\end{eqnarray}
where $\alpha=\alpha(r^n)$ for brevity.  Now, as shown above, the the
drift given by Eq.~(\ref{eq:drift2}) originates from the dissipation
term in Eq.~(\ref{eq:verlet1}), while the noise term in that equation
has a zero mean. The Stratonovich convention interprets the friction
term in essentially the same manner as our convention, but it also
incorrectly assumes that $\bar{\alpha}_t=\bar{\alpha}_r$. While this
seems like a minor difference, it directly causes the noise term to
produce undesirable drift. Specifically, in the Stratonovich picture
the noise term in Eq.~(\ref{eq:verlet1}) is given by
$\sqrt{2\bar{\alpha}_r k_BT dt}\sigma^{n+1}$, the mean value of which
should be added to the middle part of Eq.~(\ref{eq:drift1}). Using the
expansion $\bar{\alpha}_r\simeq \alpha+\alpha^\prime\Delta r/2$ and
the relationship expressed by Eq.~(\ref{eq:sigmar}), we find that the
noise term in the Stratonovich convention generates a drift of size
$\langle \Delta r\rangle =+(\alpha^\prime/\alpha)(k_BT/m)(dt^2/4)$,
which eliminates half of the drift from the dissipation term. This
error in the Stratonovich convention can be corrected by simply
shifting the value of $r^{n+1}$ computed by the Langevin integrator
Eqs.~(\ref{pos-equation})-(\ref{vel-equation}) by
$-(\alpha^\prime/\alpha)(k_BT/m)(dt^2/4)$. The corrected-Stratonovich
convention (which, as demonstrated in Fig.~\ref{fig:results2}, also
produces a fairly uniform distribution) corresponds to the overdamped
limit of the Langevin equation when reached by adiabatic elimination
of the velocity \cite{sancho}. In the isothermal convention,
$\alpha(r^{n+1})\simeq \alpha+ \alpha^\prime\Delta r$ is used in both
the dissipation and noise terms. When compared to the Stratonovich
case, the difference is that $\alpha(r^{n+1})-\alpha(r^{n})\simeq
2[\bar{\alpha}_r-\alpha(r^{n})]$. This implies that in the isothermal
convention, the dissipation term produces a drift which is twice
larger than necessary, and the noise term eliminates half of this
drift to bring it to just the right value. In other words, the
isothermal convention benefits from the cancellation of two
errors. The new convention, presented here, produces the correct drift
from the dissipation term, and no drift (as should be) from the
noise. We argue that this feature makes it the most physical approach,
despite of the fact that all the three methods presented in
Fig.~\ref{fig:results2} produce distributions that closely match the
uniform equilibrium distribution.

\begin{acknowledgements}
OF acknowledges Tamir Kamai for discussions on the
It\^{o}-Stratonovich dilemma.  This project was supported in part by
the US Department of Energy Project \# DE-NE0000536 000.
\end{acknowledgements}


\begin{thebibliography}{99}
\bibitem{footnote2} The spurious force cannot be introduced into the
Langevin equation as a real force since, unlike real forces, it cannot
be associated with a potential energy gradient that leads to
variations in the equilibrium distribution.
\end{thebibliography}
\end{document}